\documentclass[12pt,preprint]{aastex}

\slugcomment{}

\shorttitle{Kinematics of the Broad Line Region in M81}
\shortauthors{Devereux et al.}

\begin{document}


\title{Kinematics of the Broad Line Region in M81}


\author{Nick Devereux\altaffilmark{1}}
\affil{Department of Physics, Embry-Riddle Aeronautical University,
    Prescott, AZ 86301}
\email{devereux@erau.edu}

\and

\author{Andrew Shearer}
\affil{Physics Department, National University of Ireland, Galway}
\email{andy.shearer@nuigalway.ie}


\altaffiltext{1}{Fulbright Scholar, Physics Department, National University of Ireland, Galway}


\begin{abstract}
A new model is presented which explains the origin of the broad emission lines observed in 
the LINER/Seyfert nucleus of M81 in terms of a steady state spherically 
symmetric inflow, amounting to ${\sim}$1 x 10$^{-5}$ 
M${_{\sun}}$/yr, which is sufficient
to explain the luminosity of the AGN. The emitting volume has an outer 
radius of ${\sim}$1 pc, making it the largest broad line region yet to be measured, and
it contains a total mass of ${\sim}$ 5 ${\times}$ 10$^{-2}$ M${_{\sun}}$  of dense, ${\sim}$  ${10^{8}}$  cm${^{-3}}$, ionized gas, 
leading to a very low filling factor of ${\sim}$ 5 x 10$^{-9}$. 
The fact that the BLR in M81 is so large may explain why the AGN is unable to sustain the ionization
seen there. Thus, the AGN in M81 is not simply a scaled down quasar.
 \end{abstract}





\section{Introduction}

Broad emission lines \citep{Sey43} are a defining characteristic of active galactic nuclei (AGN). 
They occur
in AGNs spanning seven decades in luminosity, from the lowest luminosity Seyferts to the
highest luminosity quasars. A variety of models have been proposed to explain the origin of the broad
emission lines, but they 
all have one feature in common; that the lines are formed in the 
dense, ${\sim}$ ${10^{8}}$ to ${10^{9}}$ cm${^{-3}}$, and fast moving, ${\sim}$ ${10^{4}}$ km/s, gas, close to a massive black hole (MBH) \citep{Res77}. The models differ only in how the 
gas achieves it's high velocity, be it through inflow, outflow, spinning in a disk or attached to
stars orbiting close to the AGN.  Identifying the mechanisms that produce broad emission lines is 
of paramount importance as only then will we finally be able to understand, and exploit, the 
physics of the
small and as yet, spatially unresolved broad emission line regions (BELRs) in galaxies.

The main purpose of this paper is to draw attention to an important subset of AGNs
that exhibit flat-topped broad emission line profiles which may be utilized to determine
a size for the broad line region (BLR) that heretofore has previously only been possible using reverberation mapping
techniques (\cite{Pet93}, \cite{Pet01}). 
We demonstrate the utility of this new technique by interpreting the flat-topped broad H${\alpha}$ emission line
profile of the LINER/Seyfert nucleus of M81.

At a distance of  3.6 Mpc \citep{fre01} M81
contains the nearest and best resolved low luminosity AGN (\cite{Pei81}, \cite{Fil88}). The almost face-on aspect of the host
galaxy allows a clear and unobscured view of the nucleus which has been studied across the electromagnetic spectrum ( \cite{Pei81}, \cite{Hol96}, \cite{Bow96}, \cite{Bie00}, \cite{Lap04}).
M81 harbors a black hole (BH) with a mass, M${_{BH}}$ = 
7 ${\times}$ 10${^{7}}$ M${_{\sun}}$, deduced from a 
two dimensional velocity map obtained with STIS on the HST \citep{Dev03}. The AGN is identified with a 
UV point source at the nucleus which has a luminosity at 1500 {\AA},  L$_{1500}$ = 5 ${\times}$ 10${^{5}}$ L${_{\sun}}$ \citep{Dev97} and an X-ray (0.5 - 2.4 keV) luminosity of 5 ${\times}$ 10${^{6}}$ L${_{\sun}}$ \citep{Lap04}.

Our paper is based on the premise that the broad emission lines (BELs) are produced in a kinematically distinct 
and hence physically distinct region centered on the BH. Therefore, the broad emission line region (BELR) has a well defined inner radius and outer radius.
Since the gravitational field strength is
already known in M81, the relationship between velocity and radius may be established, given a 
kinematic model for the BELR gas. Thus, in principle, one can determine the inner and outer radii
of the BELR in M81 by modeling the broad emission line profiles.

The layout of our paper is as follows. In section 3, we evaluate the origin of the BELs seen in
M81 in the context of
viable models for the production of BELRs in AGNs, namely inflow, outflow, rotating gas disks 
and the atmospheres of stars illuminated by the 
AGN. In section 4, we discuss the viable models in the context of various  
constraints and conclude, 
in section 5, that inflow provides the best explanation for the BELR in M81. 
We begin, however, with a review
of the broad H${\alpha}$ emission line  observed in M81.

\section{The Broad H${\alpha}$ Emission Line Profile in M81}


The broad H${\alpha}$ emission line, measured by STIS aboard the HST, has been presented 
previously \citep{Dev03} and is shown in Figure 1 for convenience. 
As the  figure illustrates, the profile exhibits a rather unusual ${^{\prime}}$flat-top${^{\prime}}$,  that is rarely seen in AGNs.  
Such profiles may be
produced by radiation from a spherically symmetric shell of gas in radial motion as noted 
previously by \citet{cap80}. Single peaked emission line profiles may also be produced
by accretion disks \citep{Che89} and bipolar flows \citep{Zhe90}.

As reported in \cite{Dev03}, the entire 
broad H${\alpha}$ line is redshifted by ${\sim}$ 7 {\AA} (${\sim}$ 300 km/s) with respect to the 
narrow H${\alpha}$ line. The broad line is very symmetrical and so the redshift can not be attributed to skewness in the line shape. Similar redshifts have been noted for broad emission lines in other AGNs
\citep{Era03}. We presently have no explanation for this phenomenon, and we overlook it
for now, with the consequence that the velocities reported in this paper are 
relative to the center of the respective emission lines.

Slight irregularities on the redward side of the profile, near the peak,
coincide with the narrow components of the 6563 {\AA} H${\alpha}$ line and 
the brighter of the two [NII] lines, each of which had to be subtracted to isolate the broad H${\alpha}$
component, as described in more detail by \cite{Dev03}. The blue side of the profile, however, is less susceptible to the subtraction procedure because it coincides with the fainter [NII] line. Interestingly, 
the H${\beta}$ line is similar in shape to the 
H${\alpha}$ emission line as noted previously by \cite{Fil88} and \cite{Bow96}. The distinctive ${^{\prime}}$flat-top${^{\prime}}$ is thus well established for M81. 
Other broad emission lines are observed in M81, as described previously
by \cite{Hol96}, but the Balmer lines are the brightest, they are presently the best resolved spectrally and hence the easiest to model.

The broad Balmer lines in M81 are apparently variable. Faint, widely separated
 ${^{\prime}}$double peaks${^{\prime}}$ did appear on the wings of the Balmer lines
as reported previously by \cite{Bow96}, but they did not appear in the more recent STIS spectrum. 
The transient features are consistent with the 
disruption of a star at the tidal radius of the MBH in M81, but the broad line itself remains
to be explained.




\section{Broad Line Region Models}

A large body of work exists addressing the origin of broad lines in luminous AGNs, but
the problem of the formation and confinement \citep{mat87}, the excitation \citep{Kwa81} and 
the kinematics \citep{Robp90} of the broad line gas remains unsolved. Interestingly, broad 
lines also occur in
low luminosity AGNs \citep{Fil85} which presents the opportunity to examine the phenomenon
from a different perspective. Furthermore, rather than attempting to explain the observed
properties of \textit{all} low luminosity AGNs, we approach the problem from the 
point of view of examining one well studied example, namely M81.

\subsection{Outflow Model}

Evidence for outflow is seen in the spectra of broad absorption line quasars (e.g. \cite{Haz84}, \cite{Ara99}). However, the outflow model can be ruled out for M81 because the diminutive luminosity generated by the AGN is simply unable to provide sufficient radiation pressure to overcome the gravitational force of the MBH as demonstrated in the following. 

The condition for a radiatively driven wind is given by, 

\begin{equation}
\kappa L/ 4 \pi c G > M 
\end{equation}

where M and L are the mass and luminosity respectively, interior to a radius r, c is the speed of light, and G the gravitational constant. For M81, this condition
is not satisfied by six orders of magnitude, if one adopts the Thompson scattering opacity 
${\kappa}$ = 0.4 cm${^{2}}$/g for a pure hydrogen gas, M${_{BH}}$ = 
7 ${\times}$ 10${^{7}}$ M${_{\sun}}$, and  a combined UV and X-ray luminosity of 5 ${\times}$ 10${^{6}}$ L${_{\sun}}$. The disparity is simply too large to overcome even by invoking line opacities, as line driven winds are viable only for objects that radiate close to the Eddington luminosity limit (\cite{King03}, \cite{shl85}). A short, $1.7 \times 10^{-2}$ pc, one sided radio jet \citep{Bie00} is seen
to emerge from the nucleus of M81, but this is an unlikely source for the observed
broad line emission as that outflow consists of a relativistic plasma.

\subsection{Stellar Atmospheres Illuminated by the Central AGN Model}

We are also able to rule out another contending theory for the production of
broad emission lines in M81, namely ionization of the extended mass loss envelopes of red giant stars orbiting close to the AGN
(eg. \cite{sco88}, \cite{ale97}). Stellar winds provide a natural explanation for the confinement and replenishment of the so called ${^{\prime}}$broad line clouds${^{\prime}}$, which is otherwise a major problem \citep{mat87}. However, such a model is unlikely to be the explanation for the BELR in M81, because the intensity of ionizing photons at the tidal radius is
insufficient to penetrate the denser, ${\sim}$ ${10^{8}}$ cm${^{-3}}$, region of the mass loss wind, as shown 
in the following. 

The tidal radius, R${_{tidal}}$, for stars in synchronous rotation around a BH is given by
\begin{equation}
R{_{tidal}} = R{_{star}} [ M{_{BH}}/ M{_{star}} ]{^{1/3}}
\end{equation}

For a 1 M${_{\sun}}$ star of radius ${10^{13}}$ cm, in the
vicinity of a 7 ${\times}$ 10${^{7}}$ M${_{\sun}}$ BH, the tidal radius corresponds to  4 x ${10^{15}}$ cm.

Following \cite{sco88}, and assuming plane parallel geometry, the ionizing photon flux at the tidal radius is related to the penetration depth, d, in the inverted Stromgren sphere of the stellar wind by

\begin{equation}
 \int^\infty_d \alpha_B n^2_H dr =  \int^{\nu_{max}} _{\nu_{o}} [  L  / (4 \pi R{_{tidal}}^2 h\nu) ] d\nu 
\end{equation}

where ${\alpha_B}$ = 2.6 x ${10^{-13}}$ cm${^{-3}}$ s${^{-1}}$, is the Case B recombination coefficient, and h${\nu}$ is the energy of an ionizing photon. 

For a power law of spectral index ${\alpha}$,

\begin{equation}
L = L_{o} (\nu/  \nu_o)^{-\alpha}
\end{equation}

the right hand side of equation 3 integrates to

\begin{equation}
 \int^\infty_d \alpha_B n^2_H dr = L_{o}  \nu_o^{\alpha} [  \nu_o^{-\alpha} -  \nu_{max}^{-\alpha}] /  (h \alpha  4 \pi R{_{tidal}}^2)
\end{equation}

The hydrogen number density, n${_H}$, is related to the stellar mass loss rate, $\dot{M}$, 
and the wind velocity, v${_W}$, via the
equation of continuity;

\begin{equation}
n_H(r) = \dot{M} / (4 \pi r^2 v_W m_H)
\end{equation}

Combining equations 3 and 6, and evaluating the left hand integral, one finds,

\begin{equation}
d =  [R{_{tidal}^2}    \alpha_B h \alpha \dot{M}^2 /(4 \pi  L_{o}  \nu_o^{\alpha} [  \nu_o^{-\alpha} -  \nu_{max}^{-\alpha}]  v_W^2 m_H^2)]^{1/3}
\end{equation}

Substituting equation 7 into equation 6, and adopting ${\alpha}$ = 2,
h${\nu_{o}}$ = 13.6 eV and h${\nu_{max}}$ = 600 eV \citep{Hol96}, one obtains for the number density in the wind, \textit{n}, at the penetration depth, \textit{d}, of the ionizing photons;

\begin{equation}
n(d) = 63 \times10^6 [L_{1500}/10^5 L{\sun}]^{2/3} [R{_{tidal}}/10^{15} cm]^{-4/3} [v_W/10 km/s]^{1/3} [\dot{M}/ 10^{-5} M{\sun}/yr]^{-1/3}  cm^{-3}
\end{equation}

Substituting representative values for  the wind velocity v${_W}$ = 10 km/s, 
and stellar mass loss rates of $\dot{M}$ = 10${^{-5}}$  M${_{\sun}}$/yr \citep{sco88} and 
$\dot{M}$ = 10${^{-6}}$  M${_{\sun}}$/yr
 \citep{ale97}, one finds for M81 that representative densities in the wind at the penetration depth of 
 the ionizing photons are 30 x ${10^{6}}$  cm${^{-3}}$ and 63 x ${10^{6}}$ cm${^{-3}}$,
 respectively. Thus, the density where most of the H${\alpha}$ emission is produced is 
expected to be significantly lower than the ${\sim}$  ${10^{8}}$  cm${^{-3}}$ required for the broad line 
region of M81 to not produce broad forbidden lines \citep{Hol96}. Furthermore, the densities 
calculated above are for stars as 
close to the UV source 
as the tidal radius permits, and assume no absorption of ionizing photons inside that volume. Obviously, the penetration depths and densities will
be lower for the majority of stars which are further away. Plus, it is quite likely 
that there
is gas inside the tidal radius that would attenuate the ionizing flux, thereby exacerbating the problem. 
An additional argument against stars is that, for M81, the emitting envelopes would occupy
such a narrow range of orbital velocities that they would be unable to reproduce the observed broad
line profile. We therefore conclude that the BELR in
M81 is not produced by the extended
mass loss envelopes of red giant stars orbiting close to the AGN.

\subsection{Accretion Disk Model}

The next contender for the origin of the broad emission lines
is a rotating ${^{\prime}}$accretion disk${^{\prime}}$. 
While there has 
been a novel proposal \citep{mur97}
to generate a single peaked profile using a disk wind, we are reluctant to adopt it
as there is no \textit{a priori} evidence that there is an outflowing wind in M81.
Instead, the  
model used here involves a relativistic accretion disk developed by \cite{Che89}, and adopted
later by \cite{Era01} to explain the single peaked broad line profile in the LINER nucleus NGC 3065. 
An illustrative fit to the broad H${\alpha}$ emission line in M81, using the relativistic disk model,
is presented in Figure 2.

The model invokes five free parameters, a dimensionless inner radius, ${\zeta}_{1}$, and outer 
radius, ${\zeta}_{2}$, 
an inclination angle, i,
an emissivity law, q, and a velocity dispersion for the gas, ${\sigma}$. As described previously by \cite{Era01},
the two emission peaks; a distinguishing characteristic of disk models, 
may be merged into a single peak, and the profile wings made symmetric, if the disk has a large
outer radius and is inclined to the line of sight. The model also incorporates a 300 km/s velocity dispersion to widen the two velocity peaks
so that they merge into a reasonable representation of the flat-top profile
seen in M81. Additionally, the model incorporates a q=2 emissivity law which 
accentuates emission from the 
inner regions of the disk. 
For a BH with a mass the same as measured for M81, 
the outer radius of the disk model, illustrated in Figure 2, is large, corresponding to ${\sim}$ 0.8 pc.

The success of the model in fitting the profile is dependent on the disk inclination. If the disk
is nearly face-on, as observed for the narrow line gas \citep{Dev03}, then the inner radius
of the disk has to be small to achieve the necessary range of velocities at zero intensity. Consequently, 
gravitational and
transverse redshifts become discernable, with the result that the red wing becomes asymmetric,
somewhat extended and unlike the observed profile. 
However, a disk inclination of 50 degrees is able to reproduce the shape
of the M81 profile quite well as Figure 2 illustrates. Such a high inclination
for the inner disk is supported by observations of the radio jet \citep{Bie00}. 
Thus, if the broad emission line is produced by an accretion disk, then the unresolved inner part
of the disk is considerably more inclined than the larger scale disk seen in 
ground based and space based images.

\subsection{Evidence for Inflowing Gas In M81}

As alluded to in the introduction, a steady state spherical inflow is also able to reproduce the 
broad H${\alpha}$ emission line profile 
observed for M81, and with the minimum of 
assumptions. The relationship between velocity and radius is determined by the mass distribution, 
\textit{M(r)},
to be

\begin{equation}
v(r) = -\sqrt{2 G M(r) / r}
\end{equation}

where \textit{M(r)} includes a point mass, representing the black
hole, embedded in the center of an extended star cluster, as described previously by \cite{Dev03}. The observed, radial component, of the velocity for each particle, \textit{i}, is given by

\begin{equation}
v_{i} = v(r) cos \Theta
\end{equation}

where ${\Theta}$ is the angle between the radial velocity vector and the line of sight. 

A 
broad line profile, ${\Phi(v)}$, with the characteristics of the one observed in M81 may be produced by
generating a histogram of velocities for a system of particles distributed randomly within a 
spherical volume bounded by an inner radius, ${r_{inner}}$, an outer radius, ${r_{outer}}$ and a
cloud number density distribution, N(r), such that,

\begin{equation}
\Phi (v) =  \sum_{r_{inner}}^{r_{outer}} N(r) v_{i}
\end{equation}

For a steady state flow, where $\dot{M}$ ${\neq}$ \textit{f(r)}, the cloud number density distribution is determined by mass conservation to be,

\begin{equation}
N(r) \propto r{^{-3/2}}
\end{equation}

As described previously by \cite{cap80}, the velocity width of the ${^{\prime}}$flat-top${^{\prime}}$ is determined by the outer radius, and the width at zero intensity is determined by the inner radius. Thus, given a density
distribution and a velocity law, one can use the profile shape to determine the physical size of the emitting region. Values that reproduce
the best representation of the broad H${\alpha}$ emission line profile, shown in Figure 1, 
correspond to an
outer radius, ${r_{outer}}$ = 1 pc, and an inner radius,  ${r_{inner}}$ = 0.2 pc. 

There are other broad lines seen in M81, but
they are fainter and were acquired with lower spectral resolution. However, two of the better examples
appear in the UV;
Ly${\alpha}$ at 1216  {\AA} and Mg II at 2800  {\AA}. They are reproduced in Figures 3 and 4 respectively. 
According to \citet{Hol96}, the central depression seen in the Ly${\alpha}$ line is 
likely caused by interstellar absorption and the bright central narrow line is geocoronal emission.
They suggest that interstellar absorption is also responsible for the structure seen in the core of the 
Mg II line.
Nevertheless, the $\it{wings}$ of the Ly${\alpha}$ and Mg II profiles can be reproduced using the same inflow model and yield slightly different sizes for the emitting region. For example, the Ly${\alpha}$ emission line requires ${r_{outer}}$ = 0.7 pc, and an inner radius,  ${r_{inner}}$ = 0.1 pc. The Mg II emission line requires 
${r_{outer}}$ = 0.9 pc, and an inner radius,  ${r_{inner}}$ = 0.35 pc. These 
illustrative fits are shown in Figures 3 and 4. Overall, the sizes determined for the BELRs are similar. 
We note
that the line width at the intensity of the continuum sets an upper limit on the inner radius. This is because the line may well persist below the continuum level, even though it can not be directly measured. 

\section{Discussion}

Evidently, two plausible models; a rotating accretion disk and a spherically symmetric inflow, are able to
reproduce the shape of the broad H${\alpha}$ emission line profile observed for M81. In either
case, the velocity law dictates that the emitting region is large;  ${\sim}$ 0.7 to 1 pc in radius.
In the context of the accretion disk model, the broad H${\alpha}$ emission line presumably represents
the continuation of the larger scale H${\alpha}$ disk, seen in ground based
and space based images \citep{Dev97}, as it extends down into the unresolved nuclear 
region. On the other hand, the interpretation of the broad H${\alpha}$ emission line profile
in terms of a spherically symmetric inflow would represent a 
new phenomenon for M81. 

\subsection{Broad Line Region Size}

Perhaps the most surprising result to have emerged from our analysis is the 
large outer radius, ${\sim}$ 1 pc, inferred for BELR of M81, regardless of whether
the broad emission line is produced by an accretion disk or an inflow.
With an angular diameter of 0.1${\arcsec}$, 
the BLR of M81 is on the brink of being resolved with HST. The large outer diameter inferred for the BLR may explain why a ${^{\prime}}$flat-top${^{\prime}}$
profile is observed in M81, and rarely in other AGNs. Unless the BELR is resolved, the line
profiles will be contaminated by the slower moving gas surrounding the BELR, causing
the line profiles to be peaked rather than flat. 

The size we infer for the BELR causes M81 to not conform to the correlation between
BLR size and UV luminosity established for higher luminosity AGNs (\cite{Pet01}, \cite{Pet93}).
However, as noted by \cite{Kas05}, the correlation appears to break down for low luminosity 
AGNs, with L$_{UV}\lesssim$ 10${^{6}}$ L${_{\sun}}$, which would include M81. But the
large size that we infer for the BELR of M81 does make it the largest ever measured. On the other hand, there are good reasons to believe the sizes of broad line regions may have
been underestimated, previously, using the reverberation mapping technique, 
which is most sensitive to measuring the inner radius, closest to the AGN (\cite{Ede88}, \cite{Rob90}).

\subsection{Broad Line Region Ionization}

The size we infer for the BELR in M81 is orders of magnitude larger than 
the range of values cited by \cite{Hol96}
but their estimate is based on the \textit{ excitation parameter} which links the excitation of the broad 
emission lines to the central UV source.
However, as noted previously by \cite{Hol96} and \cite{Mao98}
the nuclear UV source in M81 produces insufficient ionizing photons to explain the luminosity
of the broad H${\alpha}$ line, by a factor of 11 according to our calculations. We find that the 
the central AGN produces 9 x ${10^{49}}$ ionizing ph/s, integrated between 13.6 and 600 eV, 
assuming a ${\nu^{-2}}$ power law, whereas excitation of the broad H${\alpha}$ emission line requires 
1 x ${10^{51}}$ ionizing ph/s. \cite{Hol96} also note that the BLR does not respond to time variable X-ray emission. In our view, this provides further evidence that the central AGN is not responsible
for ionizing the BLR gas. That the BELR in M81
is so large may explain why the central AGN is unable to sustain the ionization seen there. 
Thus, the low luminosity AGN in M81 is not simply a scaled down quasar.

The origin of the
excitation for the BELR is as enigmatic as the origin of the excitation for the
more extended narrow line region in M81, discussed previously by \cite{Dev97}. Most likely, the 
gas is excited, in situ, by shocks, or by hot, and as yet undetected, PAGB stars. In either case, the excitation parameter
would then be completely unrelated to the central source, and hence would
not yield a useful measure of the BELR size.

\subsection{Virial Black Hole Masses}

Previous attempts to estimate the BH mass in M81 using
the so called ${^{\prime}}$Virial Method${^{\prime}}$ (\cite{Pei81}, \cite{Fil88}, \cite{Hol96}) have 
consistently underestimated the kinematically determined mass \citep{Dev03}  by factors ranging 
from 20 to 200. For example, the recent formalism
of \cite{Gre05}, which uses the FWHM {\it and} luminosity of the broad H${\alpha}$ emission line, 
underestimates the mass of the BH in M81 by a factor of 140. On the other hand, the BH in
M81 does conform to the BH mass-bulge velocity
dispersion correlation  \citep{Fer00} as noted previously by \cite{Dev03}. This dichotomy is 
regarded as further evidence
that the BLR in M81 is very different from those studied in more luminous AGNs.

\subsection{The Mass of Ionized Gas in the BELR of M81}

The mass of emitting gas may be deduced from standard (Case B) recombination theory;

\begin{equation}
M_{emitting} =  L (H_\alpha) m_H /  n_H  {\alpha^{eff}_{H\alpha}} h \nu_{H\alpha} 
\end{equation}

Using an effective recombination 
coefficient ${\alpha^{eff}_{H\alpha}}$ =
 8.6 x 10$^{-14}$ cm${^{-3}}$ s${^{-1}}$, assuming a \textit{constant} average density n = 10$^{8}$ 
 cm${^{-3}}$, and a luminosity L (H$_{\alpha}$) = 3.7 x 10${^5}$ L${_{\sun}}$ based on the broad line flux reported in \cite{Dev03}, one finds M$_{emitting} $ = 0.05 M${_{\sun}}$. This mass is 
implausibly small for an accretion disk, but can be more easily reconciled with an inflow as shown in the 
following.

\subsection{The Filling Factor and the Inflow Rate for the BELR of M81}
 
It is straight forward to calculate the 
filling factor, ${\epsilon}$, once the dimensions of the emitting region have been established. 
For a uniform density medium occupying a spherical volume of radius r, one finds 

\begin{equation}
\epsilon = 3 L (H_\alpha)/ 4  \pi  n_H^2  {\alpha^{eff}_{H\alpha}} h \nu_{H\alpha} r^3
\end{equation}

Again, using an effective recombination 
coefficient ${\alpha^{eff}_{H\alpha}}$ =
 8.6 x 10$^{-14}$ cm${^{-3}}$ s${^{-1}}$, assuming a \textit{constant} average gas density n = 10$^{8}$ 
 cm${^{-3}}$, and a luminosity L (H$_{\alpha}$) = 3.7 x 10${^5}$ L${_{\sun}}$ based on the broad line flux reported in \cite{Dev03}, one finds ${ \epsilon}$ ${\sim}$ 5 x 10$^{-9}$ for M81.

Having established the dimensions of the emitting region and the filling factor one 
can now calculate the mass inflow rate for
the ionized gas, using the equation of continuity;

\begin{equation}
 \dot{M}  =  \epsilon 4 \pi r^2 v(r)  n_H(r) m_H
\end{equation}

The velocity at the inner radius of 0.2 pc is determined by the mass distribution to be 1764 km/s. Setting the gas density in the flow to be ${10^{8}}$ cm${^{-3}}$ one obtains a mass inflow rate,  $\dot{M}$ = 1 x 10$^{-5}$ M${_{\sun}}$/yr. If the gas density is higher, the mass inflow rate will decrease. Conversely, if only a fraction of the inflowing gas is ionized, then the {\it total} mass
inflow rate will, of course, be higher.

\subsection{The Luminosity of the AGN}
 
A variety of models have been considered for the production of the broad H${\alpha}$ emission line
in M81, but by far the most straight forward model, that involves the minimum of assumptions, is the
inflow model. Accretion onto a compact massive object has long been suspected as the origin for the luminosity of the AGN in M81 \citep{Pei81} and recognizing the shape of the broad line profile may be the
signature of an inflow permits us to corroborate this idea. According to the virial theorem, the maximum
power, P, produced from the conversion of gravitational potential energy as material falls from a distance of 1pc, which for the purposes of this calculation is effectively at infinity, onto the event horizon of a MBH, can be expressed as

\begin{equation}
P = \dot{M} c^2 / 4
\end{equation}

A lower limit to the mass inflow rate that we infer from the ionized gas, 10$^{-5}$ 
M${_{\sun}}$/yr, sets a lower limit of  4 x 10$^{7}$ L${_{\sun}}$ for the power, P, which, 
without explaining the
details, is sufficient to account for both the UV and X-ray luminosity of the AGN with a conversion efficiency of  
power into radiative luminosity of ${\leq}$ 14\%. 
Interestingly, the mass inflow rate that we infer is also sufficient  to sustain the radio jet 
according to \cite{Bie00}.

\subsection{The Origin of the Broad Line Gas}
 
If the broad emission lines are produced by an inflow, then one wonders how the incredibly small mass of gas in the BELR, measured to be 5 x 10${^{-2}}$ M${_{\sun}}$, is distributed within the, by comparison, rather large emitting volume ${\sim}$1 pc in radius.
One can get an approximate estimate based on the  ${^{\prime}}$noise${^{\prime}}$ in the profile, as first suggested by \cite{Cap81}. The fact that the observed broad line profile is smoother than the model profile, which was generated using 75,000 points (see Figure 1) suggests that, on average, the mass of each ${^{\prime}}$broad line cloud${^{\prime}}$ is $ \lesssim$ 10${^{-6}}$ M${_{\sun}}$. We surmise
that representative masses for broad line clouds ${\sim}$ 10${^{-9}}$ M${_{\sun}}$, cited previously by \cite{Ale94}, are not at all unreasonable given that the observed profile in M81 is, in fact, so smooth.

It is not surprising, given the small mass of gas and the very low volume filling factor, that the 
{\it internal} extinction is negligible to both the UV source \citep{Hol96} and the gas responsible for the broad 
Balmer lines (\cite{Fil88}, \cite{Bow96}). The fact that the H${\alpha}$ profile is symmetric provides further evidence that the extinction to the BELR is
inconsequential; apparently  a common feature among AGNs (eg. \cite{Col88}).

We speculate that mass loss from a spherically symmetric distribution of late-type bulge stars may provide a more
than adequate supply for the in-falling gas. Any gas for which the vector cross product

\begin{equation}
r {\times} p = 0
\end{equation}

where r is the radius vector and p is the linear momentum of the gas, will fall into the nucleus.
Calculating how much gas satisfies this identity and furthermore, how it becomes so compressed, is beyond the scope of the present paper, but
we note that only a small mass inflow rate, ${\sim}$ 10$^{-5}$ M${_{\sun}}$/yr, of dense,  ${10^{8}}$ cm${^{-3}}$, gas is required 
to explain the luminosity of the AGN.

\section{Conclusions}
 
We have demonstrated a new technique that has allowed us to determine the size of the broad line region (BLR)  in M81 by modeling the shape of the broad line profile using a kinematic model for the broad emission line gas. Our principle conclusion is that the BLR is large, ${\sim}$ 2 pc in diameter, regardless
of whether the velocity field is represented by an accretion disk or a spherically symmetric inflow. 

The fact that the BLR in M81 is so large may explain why the AGN is unable to sustain the ionization
seen there. We therefore conclude that the gas responsible for the broad emission lines
seen in M81 must be ionized in-situ, either by shocks, hot stars, or a combination of the two. Thus,
the AGN in M81 is not simply a scaled down quasar.

How the gas becomes so compressed as to prevent the formation of
broad forbidden lines remains a mystery. Nevertheless, if the gas density really is high, 
${\sim}$ 10$^{8}$  cm${^{-3}}$, then the broad H${\alpha}$ emission line observed in the LINER/Seyfert nucleus of M81 is most easily understood in terms of a steady state spherically symmetric inflow, amounting to ${\geq}$ 1 x 10$^{-5}$ M${_{\sun}}$/yr, which is sufficient to sustain the luminosity
of the AGN and the radio jet. We further speculate, based on the smoothness of the broad emission line profile, that the individual regions of dense ionized gas have masses 
$ \lesssim$ 10${^{-6}}$ M${_{\sun}}$. Alternatively, if the gas density 
is significantly lower, then the total mass of gas increases, making the interpretation of
the broad emission lines in terms of an accretion disk more amenable, but then explaining the absence of broad forbidden lines becomes a 
problem. Thus, the gas density remains pivotal in understanding the origin of the broad
emission lines seen in M81.




\acknowledgments
The authors thank Dr. Luis Ho for providing digitized spectra, two of which were reproduced here. 
The authors also thank Prof. Mike Eracleous for his generous help with the relativistic accretion disk
modeling. N.D. gratefully acknowledges the support of the Fulbright Commission and the hospitality of the National University of Ireland, Galway, where this work
was completed. The authors
also gratefully acknowledge the support of the HEA funded Cosmogrid programme, and the referee for 
prompting us to consider the accretion disk model more carefully.
This research has made extensive use of NASA's Astrophysics Data System. 



{\it Facilities:}  \facility{HST (STIS)}

\clearpage



\begin{figure}
\epsscale{1.0}
\begin{center}
\plotone{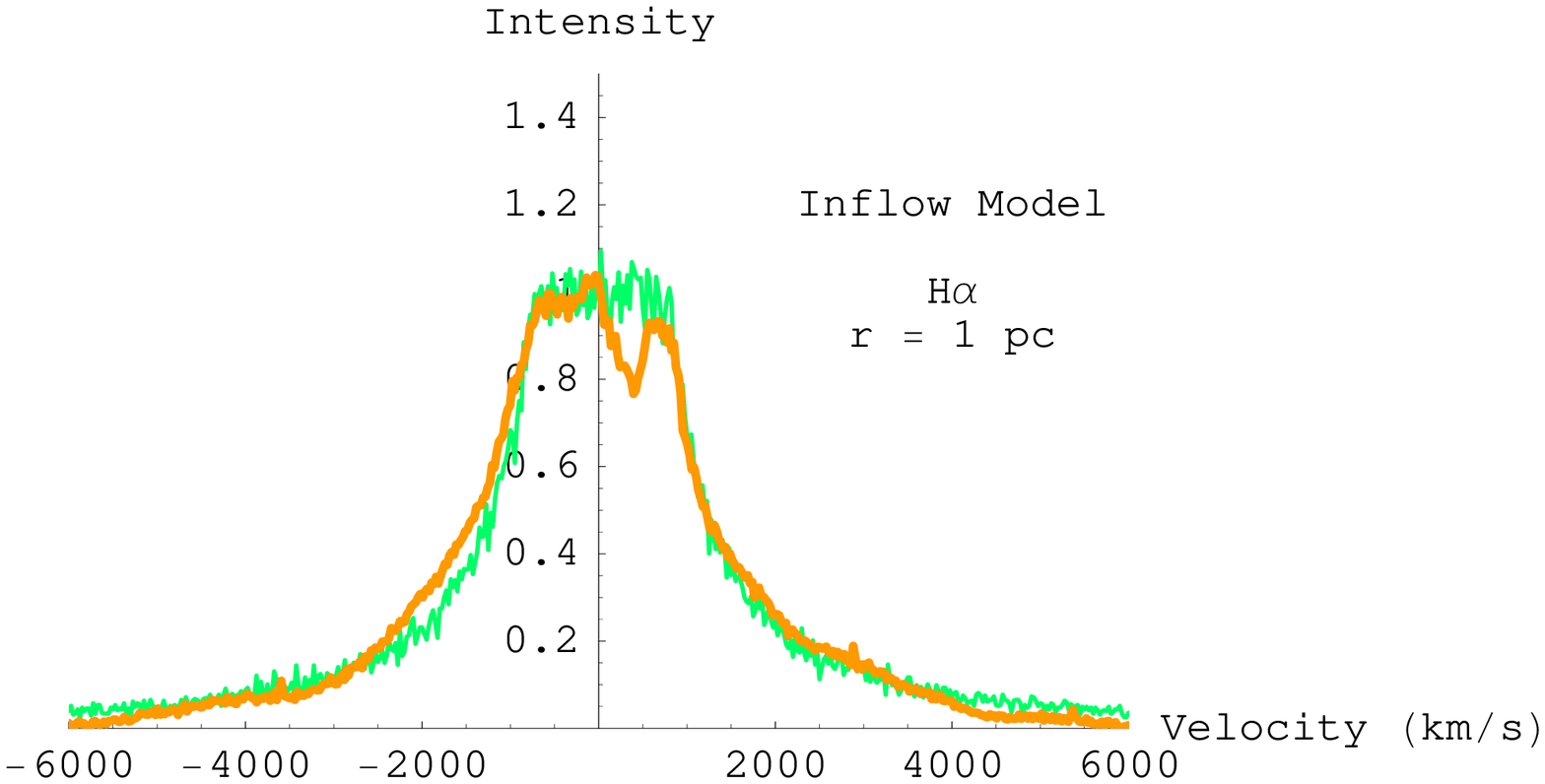}
\caption{{\bf Broad H${\alpha}$ emission line in M81. The observed profile
\citep{Dev03} is shown in orange and the inflow model is shown in green. The 
inflow was generated using 75,000 randomly distributed points in free-fall from
an outer radius of 1 pc (see text for details). 
The intensity of both profiles has been normalized 
to the intensity at zero velocity.}}
\label{default}
\end{center}
\end{figure}

\begin{figure}
\epsscale{1.0}
\begin{center}
\plotone{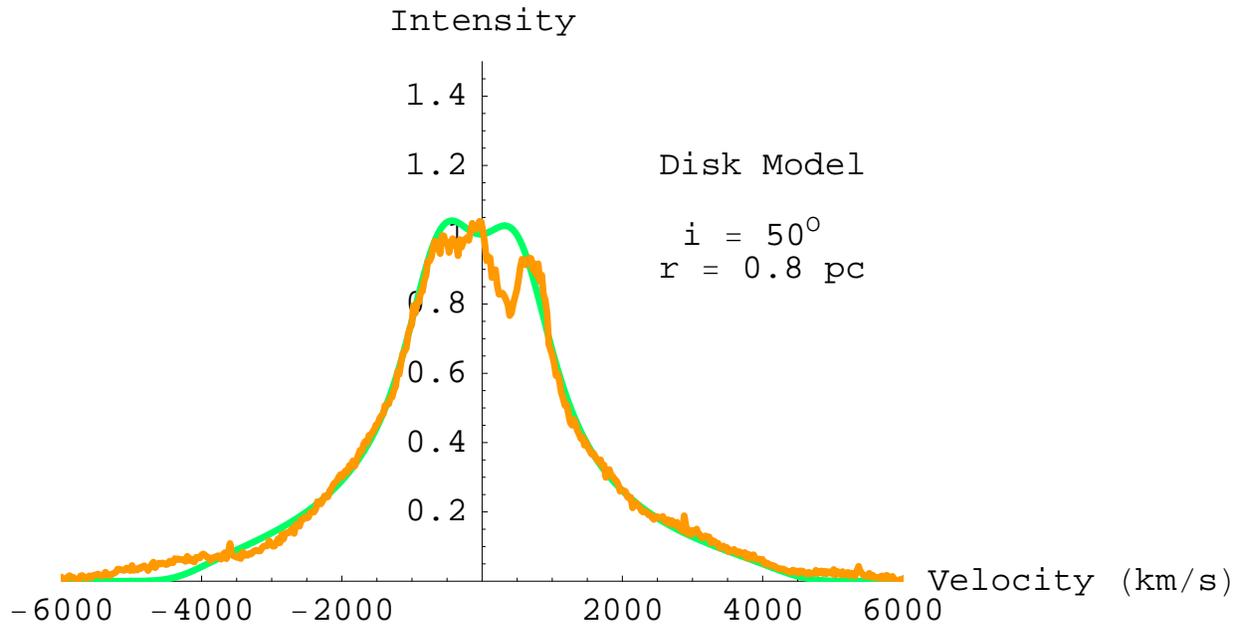}
\caption{{\bf Observed Broad H${\alpha}$ emission line in M81, shown in orange, with the disk model shown in green. The disk is 0.8 pc in radius and is inclined by 50 degrees to the line of sight (see text
for details). The intensity of both profiles has been normalized to the intensity at zero velocity.}}
\label{default}
\end{center}
\end{figure}

\begin{figure}
\epsscale{1.0}
\begin{center}
\plotone{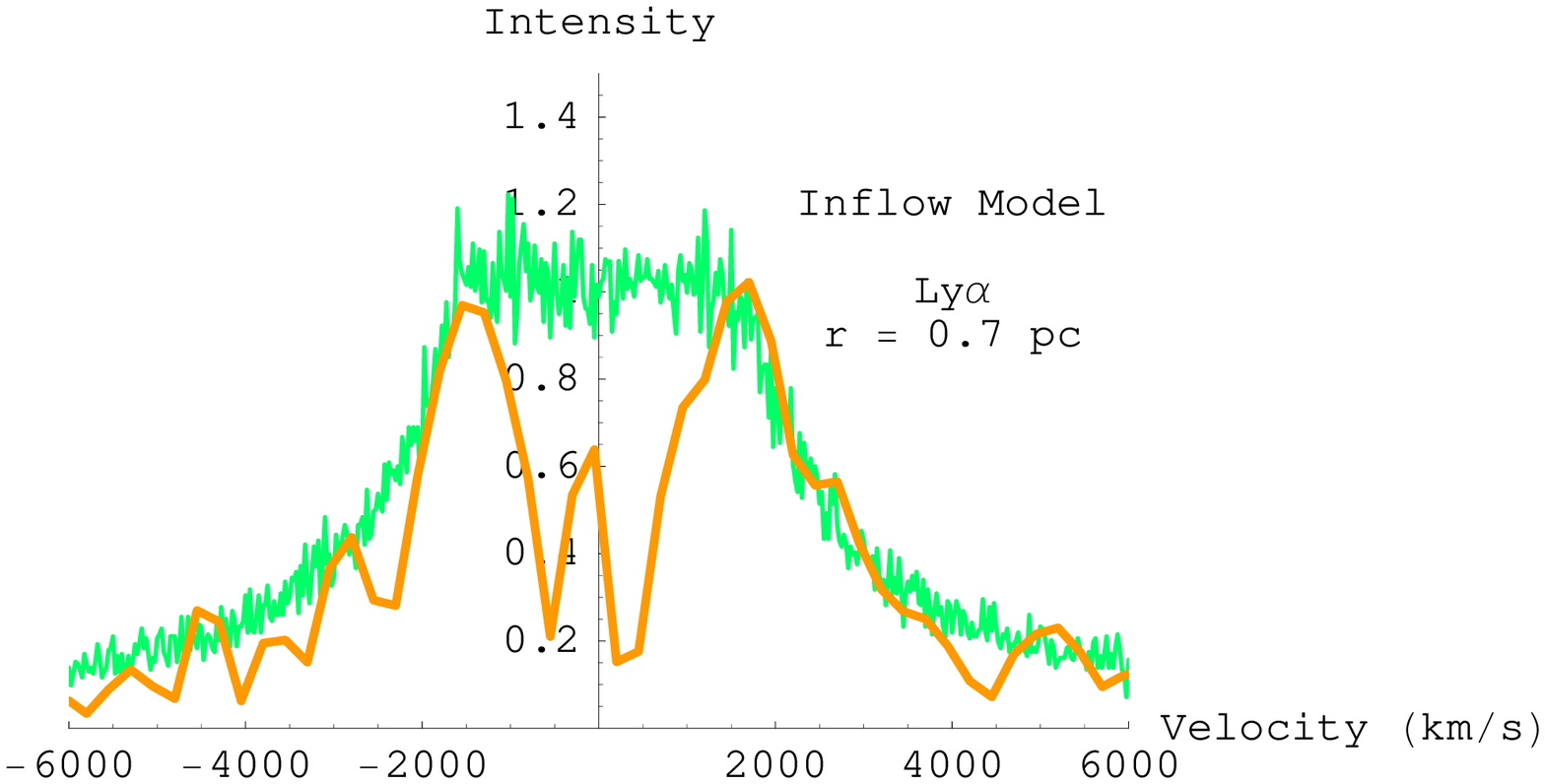}
\caption{{\bf Broad Ly${\alpha}$ emission line in M81. The observed profile 
\citep{Hol96} is shown in orange. The model profile, shown in green, represents an inflow (see text for details). The intensity of the model profile has been normalized 
to the intensity at zero velocity.}}
\label{default}
\end{center}
\end{figure}

\begin{figure}
\epsscale{1.0}
\begin{center}
\plotone{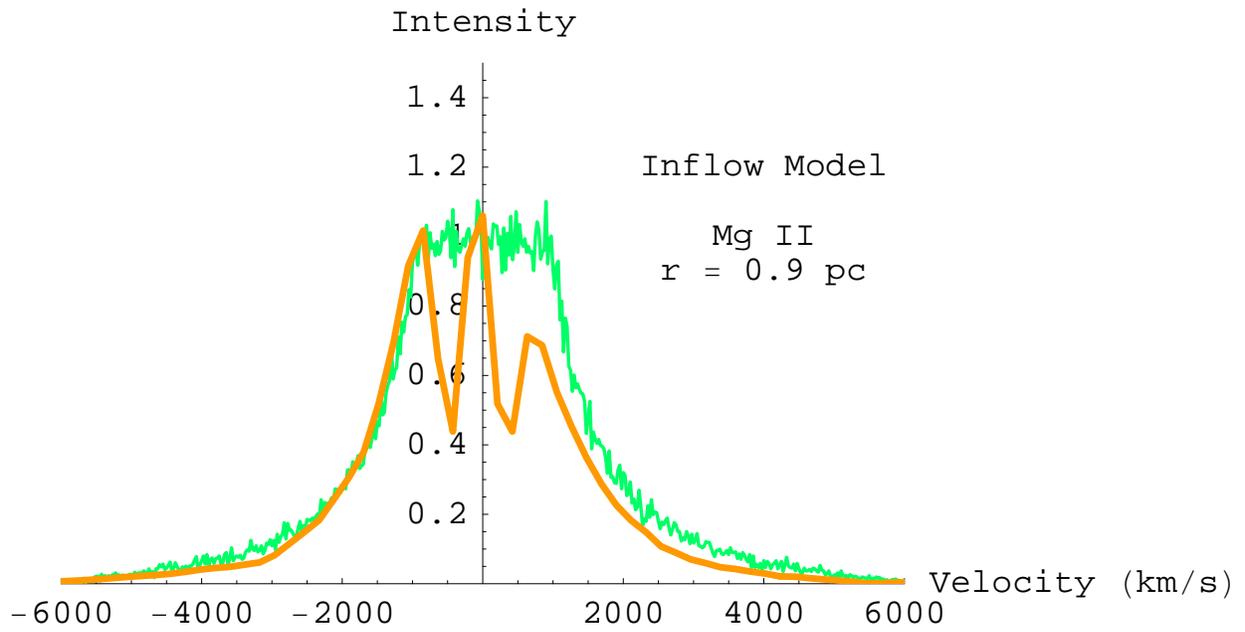}
\caption{{\bf Broad Mg II emission line in M81. The observed profile 
\citep{Hol96} is shown in orange. The model profile, shown in green, represents an inflow (see text for details). The intensity of both profiles has been normalized 
to the intensity at zero velocity.}}
\label{default}
\end{center}
\end{figure}


\begin{thebibliography}{}

\bibitem[Alexander \& Netzer (1994)]{Ale94} Alexander, T., \& Netzer, H., 1994, \mnras, 270, 781
\bibitem[Alexander \& Netzer (1997)]{ale97} Alexander, T., \& Netzer, H., 1997, \mnras, 284, 967
\bibitem[Arav et al. (1999)]{Ara99} Arav, N; Becker, R. H., Laurent-Muehleisen, S. A., Gregg, M. D.;,White, R. L., Brotherton, M. S., de Kool, M., 1999, ApJ 524, 566
\bibitem[Bietenholz, Bartel \& Rupin (2000)]{Bie00} Bietenholz, M.F.,  Bartel, M.P., 2000, \apj, 532, 895
\bibitem[Bower et al. (1996)]{Bow96} Bower, G.A., Wilson, A.S., Heckman, T.M., \&
Richstone, D.O.,1996, \aj, 111, 1901
\bibitem[Capriotti, Foltz \& Byard (1980)]{cap80} Capriotti, E., Foltz, C.,  
    \&   Byard, P., 1980, \apj, 241, 903
\bibitem[Capriotti, Foltz \& Byard (1981)]{Cap81} Capriotti, E., Foltz, C.,  
    \&   Byard, P., 1980, \apj, 245, 396
 \bibitem[Chen \& Halpern (1989)]{Che89} Chen, K., \& Halpern, J.P., 1989, \apj, 344, 115   
 \bibitem[Collin-Souffrin \& Lasota (1988)]{Col88} Collin,-Souffrin, S., \& Lasota, J.P, 1988, \pasp 100, 1041
\bibitem[Devereux, Ford \& Jacoby (1997)]{Dev97} Devereux, N., Ford, H.,  
    \&   Jacoby, G., 1997, \apjl, 481, 71
\bibitem[Devereux et al.(2003)]{Dev03} Devereux, N., Ford, H., Tsvetanov, Z., 
    \&   Jacoby, G., 2003, \apj, 125, 1226  
\bibitem[Edelson \& Krolik (1988)]{Ede88} Edelson, R.A.,\& Krolik, J.H., 1988, \apj, 333, 646
\bibitem[Eracleous \& Halpern (2001)]{Era01} Eracleous, M., \& Halpern J.P., 2001, \apj, 554, 2406
\bibitem[Eracleous \& Halpern (2003)]{Era03} Eracleous, M., \& Halpern J.P., 2003, \apj, 599, 886
\bibitem[Ferrarese \& Merritt (2000)]{Fer00} Ferrarese, L., \& Merritt, D., 2000, \apj, 539, L9
\bibitem[Filippenko \& Sargent (1985)]{Fil85} Filippenko, A.V., \& Sargent, W.L.W., 1985, \apjs, 57, 503
\bibitem[Filippenko \& Sargent (1988)]{Fil88} Filippenko, A.V., \& Sargent, W.L.W., 1988, \apj, 324, 134
\bibitem[Freedman et al. (2001)]{fre01} Freedman, W.L., et al. 2001, \apj, 553, 47
\bibitem[Greene \& Ho (2005)]{Gre05} Greene, J.E., \& Ho, L.C., 2005, \apj, 630, 122
\bibitem[Hazard et al. (1984)]{Haz84} Hazard, C.; Morton, D. C.; Terlevich, R.; McMahon, R., 1984, \apj, 282, 33
\bibitem[Ho, Filippenko \& Sargent (1996)]{Hol96} Ho, L.C., Filippenko, A.V., \&
      Sargent, W.L.W., 1996, \apj, 462, 183
\bibitem[Kaspi, et al. (2005)]{Kas05} Kaspi, S., Maoz, D., Netzer, H.,Peterson, B.M.,
Vestergaard, M., \& Jannuzi, B.T., 2005, \apj, 629, 61
\bibitem[King \& Pounds (2003)]{King03} King, A., \& Pounds, K., 1981, \mnras, 347, K657
\bibitem[Kwan \& Krolik(1981)]{Kwa81} Kwan, J., \& Krolik, J.H., 1981, \apj, 250, 478
\bibitem[La Parola et al. (2004)]{Lap04} La Parola, V., Fabbiano, G., Elvis, M., Nicastro, F., 
Kim, D.W., \& Peres, G., 2004, \apj, 601, 831
\bibitem[Maoz et al. (1998)]{Mao98} Maoz, D., Koratkar, A., Shields, J.C., Ho, L.C., 
Filippenko, A.V., \& Sternberg, A., 1998, \aj, 116, 55
\bibitem[Mathews \& Ferland (1987)]{mat87} Mathews, W.G., \& Ferland, G.J., 1987, \apj, 323, 456
\bibitem[Murry \& Chiang (1997)]{mur97} Murry, N., \& Chiang, J., 1997, \apj, 474, 91
\bibitem[Peimbert \& Torres-Peimbert (1981)]{Pei81} Peimbert, M., \& Torres-Peimbert, S. 1981, \apj, 245, 845 
\bibitem[Peterson (1993)]{Pet93} Peterson, B., 1993, \pasp, 105, 247
\bibitem[Peterson (2001)]{Pet01} Peterson, B., 2001, in Advanced Lectures on the Starburst-AGN Connection,    ed. I. Aretxaga, D. Kunth, \&  R. Mujica (Singapore: World Sci.), 3               
\bibitem[Rees(1977)]{Res77} Rees, M.J., 1977, \qjras, 18, 429
\bibitem[Robinson \& Perez (1990)]{Rob90} Robinson, A., \& Perez, E.,  \mnras, 244, 138
\bibitem[Robinson, Perez, \& Binette(1990)]{Robp90} Robinson, A., Perez, E., Binette, L., \mnras, 246, 349
\bibitem[Scoville \& Norman (1988)]{sco88} Scoville, N., \& Norman, C., 1988, \apj, 332, 163
\bibitem[Seyfert (1943)]{Sey43} Seyfert, C.K., 1943, ApJ 97, 28
\bibitem[Shlosman, Vitello, \& Shaviv (1985)]{shl85} Shlosman, I., Vitello, P.A., 
    \& Shaviv, G., 1985, \apj, 294, 96
\bibitem[Zheng, Binette \& Sulentic (1990)]{Zhe90} Zheng, W., Binette, L.,
    \& Sulentic, J.W., 1985, \apj, 365, 115


\end{thebibliography}
\end{document}